\documentclass[conference]{IEEEtran}
\IEEEoverridecommandlockouts
\usepackage{cite}
\usepackage{amsmath,amssymb,amsfonts}
\usepackage{algorithmic}
\usepackage{graphicx}
\usepackage{textcomp}
\usepackage{xcolor}
\allowdisplaybreaks
\def\BibTeX{{\rm B\kern-.05em{\sc i\kern-.025em b}\kern-.08em
    T\kern-.1667em\lower.7ex\hbox{E}\kern-.125emX}}
\begin{document}

\title{Investigating the Impact of Electric Vehicles on the Voltage Profile of  Distribution Networks
}

\author{\IEEEauthorblockN{1\textsuperscript{st} Arash Farokhi Soofi}
\IEEEauthorblockA{\textit{Department of Electrical Engineering} \\
\textit{San Diego State University}\\
San Diego, USA \\
afarokhi@ucsd.edu}
\and
\IEEEauthorblockN{2\textsuperscript{nd} Reza Bayani}
\IEEEauthorblockA{\textit{Department of Electrical Engineering} \\
\textit{San Diego State University}\\
San Diego, USA \\
rbayani@ucsd.edu}
\and
\IEEEauthorblockN{3\textsuperscript{rd} Saeed D. Manshadi}
\IEEEauthorblockA{\textit{Department of Electrical Engineering} \\
\textit{San Diego State University}\\
San Diego, USA \\
smanshadi@sdsu.edu}
}

\maketitle

\begin{abstract}
This paper investigates the impact of high penetration of Electric Vehicles (EVs) on the distribution network in the presence of photovoltaic (PV) systems. Two models for EVs are presented and the voltage profile of buses is investigated considering both models and various penetration levels of EVs within the distribution network. The analysis is conducted by presenting an exact convex relaxed form of the full ACOPF problem of the distribution network with fixed power EVs and presenting the full ACOPF problem of the distribution network with fixed current EVs. The performance of each model is illustrated in the case studies leveraging the modified IEEE 33-bus system and considering time-of-use (TOU) pricing. Besides the sensitivity of voltage profile of buses in the distribution network on the time-of-use prices is investigated in the case studies.
\end{abstract}


\section*{Nomenclature}

\subsection*{Parameters}
\noindent \begin{tabular}{ l p{6.55cm} }

$A_{s}^{t}$ & Available power from solar generation unit $s$ at time $t$\\

$B_{(.)},G_{(.)}$  & Elements of susceptance and conductance matrices \\

$C_E$ & Time-of-use price of electricity at time t \\

$\overline{E_{e}}$, $\underline{E_{e}}$  & Max./Min. energy of \color{black} fleet of \color{black} EV $e$\\

$\overline{P_{e}^{c}}$ & Maximum charging power of \color{black}fleet of \color{black} EV $e$\\

$p_d^t$,$q_d^t$ & Real and reactive power of load $d$ at time $t$\\

$P_{e,t}^{tr}$ & Traveling power consumption of \color{black} fleet of \color{black} EV $e$ at time $t$ \\

$R_d^t$, $R_c^t$ & The ratio of discharging and charging EVs to the total EVs \color{black} in the fleet \color{black} at time $t$\\

$\overline{S}_{ij}$ & The maximum apparent power flow of distribution line $(i,j)$\\

$\overline{V}_i,\underline{V}_i$  & Maximum and Minimum voltage magnitude at bus $i$\\

$\gamma_e^c,\gamma_e^d$ & Charging and discharging efficiency of \color{black} fleet of \color{black} EVs $e$\\

\end{tabular}

\subsection*{Variables}
\noindent \begin{tabular}{ l p{6.55cm} }

$E_{e,t}$ & Energy of fleet of EVs $e$ at time $t$\\

$P_{e,t}^{c}$  & Real charging power of EV $e$ at time $t$\\

\color{black}$P_g^t$, $Q_g^t$\color{black} & Real and reactive power flow between grid $g$ and the distribution network at time $t$

\end{tabular}
\newpage
\noindent \begin{tabular}{ l p{7.5cm} }

$P_s^t$ & Real power dispatch of solar generation unit $s$ at time $t$ \\

$p_{ij}^t$ &  Real power flow between bus $i$ and $j$ at time $t$\\

$q_{ij}^t$ &  Reactive power flow between bus $i$ and $j$ at time $t$\\

\color{black}$V_i^t$ \color{black} & Voltage magnitude of bus $i$ \\

$\theta_{i}$ & Voltage angle of bus $i$ \\

$c,s$ & lifting operator terms for SOCP relaxation

\end{tabular}
\vspace{-0.6cm}
\subsection*{Sets}
\noindent \begin{tabular}{ l p{7.5cm} }






$\mathcal{D}_i$ & Set of load connected to bus $i$ \\

$\mathcal{E}_i$ & Set of the electric vehicle connected to bus $i$\\


$\mathcal{G} $ & Set of all grids connected to the distribution network\\

$\mathcal{G}_{i} $ & Set of the grid connected to the distribution network through bus $i$ \\

$\mathcal{I}_e$ & Set of the bus connected to EV $e$\\







$\mathcal{S}$ & Set of all solar generation units\\


$\mathcal{S}_i$ & Set of solar generation unit connected to bus $i$ \\

$\mathcal{T} $ & Set of time horizon \\

\end{tabular}

\vspace{-0.3cm}
\section{Introduction}
 Conventional transportation is one of the most factors that pollute the environment and it will increase $60\%$ by $2050$ in the absence of vehicle electrification \cite{outlook2017oecd}. One of the major methods to decrease the emission of conventional vehicles is the adaptation of Electric Vehicles (EVs) \cite{wikstrom2016introducing}. Governments offer incentives to utilizing EVs and invest in the infrastructure of EV system \cite{amini2016arima}. The success of substituting conventional vehicles with EVs depends on preparing enough charging stations for EVs to provide sufficient charging power for batteries of EVs \cite{abdullah2020ev}. The increasing number of EV charging stations leads to an increase in utilizing EVs. Thus, the penetration level of EVs increases. The increasing penetration level of EVs leads to a high electricity load burden on the distribution network. The high electricity load burden on the distribution network creates challenges for the distribution network \cite{farhoodnea2013power,khoshjahan2021enhancing}. In \cite{amini2014allocation}, authors investigate the reliability issues of utilizing high penetration level EVs in an unscheduled manner.
 
 One of the most important challenges of increasing the penetration level of EVs is the voltage issue. The voltage profile of buses within the distribution network depends on the loads served on each bus. EVs are considered as a load for the distribution network and they can change the voltage profile of each bus based on their penetration level. Utilities are currently seeking viable solutions to remedy the problem and a wide range of options are available. \color{black} One method to mitigate the high penetration EVs effects on the distribution networks is called smart charging. Smart charging is the coordinated scheduling of the charging time and power of EVs \cite{garcia2014plug}. The algorithms proposed to control the EV charging varies in the objective, control hierarchy \cite{cheng2018comparing,clairand2018smart}, and the constraints they take into account. However, implementing smart charging approaches requires significant infrastructure to operate, it is very complex, and the willingness of the customers is required. Another method to mitigate the voltage issues of the distribution network due to utilizing high penetration EVs is deploying home solar generation units integrated with EV chargers \cite{tran2019efficient,khoshjahan2020optimal}. However, the solar generation units generate power only during sunny hours and during this period most EVs will be in the parking mode instead of being at home and connected to the grid. In this paper, fleet of EVs modeled as fixed current and fixed power loads to illustrate the impact of serving EVs as voltage regulators. Besides, the model of solar generation units is considered in the volt/VAR optimization problem to show the incapability of voltage regulation of PVs when the penetration level of EVs is high. Besides, changes in time-of-use pricing could also alternate the demand pattern which will in turn change the voltage profile. The proposed methods will not necessarily require any upgrade to the existing distribution grid.  \color{black} There are two options for modeling the load of EVs. In the first one, the charging power of EVs is considered fixed. Thus, the variation in voltage magnitude of the bus connected to the EV doesn't affect the charging power of the EV. In \cite{soofi2021analyzing}, authors investigated the impact of of utilizing EVs with high level charging rates based on the fixed power model. \color{black} However, the impact of time-of-use pricing and utilizing EVs as fixed current loads on the voltage profile of the distribution network isn't investigated. \color{black} In the second approach to model EVs, the input current of EVs is constant. Thus, variation in voltage magnitude of the connected bus leads to variation in the charging power of EVs. To evaluate the impact of the EVs on the distribution network, in this paper the AC Optimal Power Flow (ACOPF) problem of the distribution network is presented \cite{dommel1968optimal}. Due to the complexity and non-convexity of the ACOPF problem \cite{frank2012optimal}, the Second-Order-Cone Problem (SOCP) \cite{jabr2008optimal} formulation of the ACOPF problem formulation is presented. The SOCP problem formulation is exact for radial networks \cite{soofi2020conic}. Since distribution networks are radial, the SOCP relaxation form of the ACOPF problem is exact for distribution networks.
 
 This paper aims to find the answer to these questions: \textit{Does utilizing fixed current model mitigate the voltage drop of buses of the distribution network?} \textit{What is the effect of changing electricity price on the voltage drop of buses of the distribution network?}
\vspace{-0.25cm}
\section{Problem Formulation and Solution Method}
The voltage profile of buses in the distribution network depends on the loads served on each bus. Electric Vehicles (EVs) are considered as a load for the distribution system and they can change the voltage profile of each bus based on their penetration level. There are two models for EVs. In the fixed power model, the charging power of EVs is considered fixed. Thus variation in voltage magnitude of the bus connected to the EV doesn't affect the charging power of the EV. In the fixed current model, the charging current of EVs is constant. Thus, changing the voltage magnitude of the connected bus changes the charging power of EVs. The analysis is conducted by presenting the SOCP convex relaxation form of the full ACOPF optimization problem for the fixed power model and the full ACOPF optimization problem for the fixed current model of EVs. 
\vspace{-0.3cm}
\subsection{Fixed Power Model}
The SOCP relaxed form of the ACOPF problem formulation of the distribution network with solar generation units and EVs with a fixed power model is presented in \eqref{SOCP}. The objective function presented in \eqref{SOCP_obj} is minimizing the cost of the distribution network. \color{black} The real charging power of fleet of EVs $e$ at time $t$ is limited by the multiplication of the maximum charging power of fleet of EVs $e$ and the ratio of connected EVs in the fleet as presented in \eqref{SOCP_mu_pe}. \color{black} The energy balance of each electric vehicle at time $t$ is presented in \eqref{SOCP_lambda_Eet}, \eqref{SOCP_lambda_Ee1}. The physical limits of $E_e^t$ is presented in \eqref{SOCP_mu_Ee}. The physical limit of solar generation $s$ at time $t$ is presented in \eqref{SOCP_mu_s}. The SOCP relaxed form of real and reactive nodal balance constraints are presented in \eqref{SOCP_lambda_i_p} and \eqref{SOCP_lambda_i_q}, respectively. The real and reactive power flow on line $(i,j)$ are given in \eqref{SOCP_p} and \eqref{SOCP_q}, respectively. The voltage limit of each bus is presented in \eqref{SOCP_limv}. The relationship between the SOCP lifting variables and the second-order cone relaxation of the relationship between the SOCP lifting terms are presented in \eqref{SOCP_symtric} and \eqref{SOC}, respectively. In \eqref{SOCP_lim}, the thermal limit of lines is presented. The SOCP lifting variables are introduced in \eqref{soc}.
\vspace{-0.5cm}

\begin{subequations} \label{SOCP}
\begin{alignat}{3}
&\text{min}\sum_{g\in  \mathcal{G}}^{ }\sum_{t \in \mathcal{T}}^{ }C_E^tP_{g}^t\label{SOCP_obj}\\
&\text{s.t.} \hspace{0.5cm} 0\leq P_{e,t}^{c}\leq \overline{P_{e}^{c}}R_c^t\label{SOCP_mu_pe}\\
&E_{e,t}=E_{e,t-1}-(\frac{P_{e,t}^{tr}R_d^t}{\gamma_e^d}-\gamma_e^cP_{e,t}^c) \hspace{0.2cm}  \forall{t \in \mathcal{T} \setminus 1}\label{SOCP_lambda_Eet}\\
&E_{e,t}=E_{e,t-1+|T|}-(\frac{P_{e,t}^{tr}R_d^t}{\gamma_e^d}-\gamma_e^cP_{e,t}^c)\hspace{0.2cm}    \forall{t=1}\label{SOCP_lambda_Ee1}\\
& \underline{E_{e}}\leq E_{e,t}\leq \overline{E_{e}}\label{SOCP_mu_Ee}\\
&0\leq P_{s}^t\leq A_{s}^t\label{SOCP_mu_s}\\
&\sum_{s \in \mathcal{S}_i}^{} P_s^t+\sum_{g \in \mathcal{G}_i}^{}\color{black}P_g^t\color{black} =\sum_{d \in \mathcal{D}_i}^{}P_d^t+(G_{ii}+\sum_{j\in\delta_i}^{}G_{ij})c_{ii}^t+\sum_{j \in \delta_i}^{}p_{ij}^t\nonumber\\
&\hspace{3cm}+\sum_{e\in\mathcal{E}_i}^{}P_{e,t}^c\label{SOCP_lambda_i_p}\\
&\sum_{g \in \mathcal{G}_i}^{}\color{black}Q_g^t\color{black}-\sum_{d \in \mathcal{D}_i}^{}q_d^t=-(B_{ii}+\sum_{j\in \delta_i}^{}B_{ij})c_{ii}^t+\sum_{j \in \delta_i}^{}q_{ij}^t\label{SOCP_lambda_i_q}\\
&p_{ij}^t=-G_{ij}c_{ii}^t+G_{ij}c_{ij}^t+B_{ij}s_{ij}^t\label{SOCP_p}\\
&q_{ij}^t=B_{ij}c_{ii}^t-B_{ij}c_{ij}^t+G_{ij}s_{ij}^t\label{SOCP_q}\\ 
&{\underline{V}_i}^2\leq  c_{ii}^t\leq {\overline{V}_i}^2\label{SOCP_limv}\\
&c_{ij}=c_{ji} \hspace{1cm} ,\hspace{1cm} s_{ij}=-s_{ji}\label{SOCP_symtric}\\
&    \begin{Vmatrix}
2c_{ij}^t
\\2s_{ij}^t
\\ 
c_{ii}^t-c_{jj}^t
\end{Vmatrix}\leq c_{ii}^t+c_{jj}^t\label{SOC}\\
&\sqrt{(p_{ij}^t)^2+(q_{ij}^t)^2}\leq  \overline{S}_{ij}\label{SOCP_lim}\\
&\text{Where:}\hspace{0.2cm}\left\{\begin{matrix}
c_{ii}^t:=(V_i^t)^2\\
c_{ij}^t=V_i^tV_j^t\cos{\theta_{ij}^t}\\
s_{ij}^t=V_i^tV_j^t\sin{\theta_{ij}^t}
\end{matrix}\right.\label{soc}
\end{alignat}
\end{subequations} 

The SOCP relaxation problem reformulation presented in \eqref{SOCP} is exact for radial networks \cite{soofi2020conic}. Since distribution networks are radial, the SOCP relaxation form of the ACOPF problem is exact for distribution networks. The presented problem is formulated as a convex optimization problem with SOCP constraints, solar generation units, and EVs as DERs connected to the distribution network. Thus, the problem presented in \eqref{SOCP} can be solved with off-the-shelf solvers such as Gurobi \cite{gurobi}.

\vspace{-0.25cm}
\subsection{Fixed Current Model}
The ACOPF optimization problem of the distribution network with solar generation units and fixed current EVs are presented in \eqref{ACOPF}. The objective function presented in \eqref{ACOPF_obj} is minimizing the cost of the distribution network. The upper limit of the charging power of EV is presented in \eqref{ACOPF_mu_pe}. The energy balance of each electric vehicle at time $t$ is presented in \eqref{ACOPF_lambda_Eet}, \eqref{ACOPF_lambda_Ee1}. 
The real and reactive nodal balance constraints are presented in \eqref{ACOPF_lambda_i_p}, \eqref{ACOPF_lambda_i_q}. The real and reactive power injection constraints are presented in \eqref{ACOPF_p} and \eqref{ACOPF_q}, respectively. $\theta_{ij}$ is the difference between the voltage angle of bus $i$ and bus $j$. The voltage limit of lines is presented in \eqref{ACOPF_limv}. The ACOPF optimization problem given in \eqref{ACOPF} is a non-linear optimization problem and non-linearity originate from the bi-linear terms in the nodal equations and power flow equations shown in \eqref{ACOPF_lambda_i_p}-\eqref{ACOPF_q}. The rest of constraints are the same as the constraints presented in \eqref{SOCP}. 
\vspace{-0.2cm}
\begin{subequations} \label{ACOPF}
\begin{alignat}{3}
&\underset{}{\text{min}} \sum_{g \in \mathcal{G}}^{ } \sum_{t\in  \mathcal{T}}^{ }C_{E}^t\color{black}P_g^t\color{black}\label{ACOPF_obj}\\
&\text{s.t.} \hspace{0.5cm} 0\leq I_{e}^c\leq \overline{P_{e}^{c}}R_c^t\label{ACOPF_mu_pe}\\
&E_{e,t}=E_{e,t-1}-(\frac{P_{e,t}^{tr}R_d^t}{\gamma_e^d}-\gamma_e^cI_{e,t}^c\sum_{i \in \mathcal{I}_e}^{} V_i^t) \hspace{0.2cm}  \forall{t \in \mathcal{T} \setminus 1}\label{ACOPF_lambda_Eet}\\
&E_{e,t}=E_{e,t-1+|T|}-(\frac{P_{e,t}^{tr}R_d^t}{\gamma_e^d}-\gamma_e^cI_{e,t}^c\sum_{i \in \mathcal{I}_e}^{} V_i^t)\hspace{0.2cm}    \forall{t=1}\label{ACOPF_lambda_Ee1}\\
&\sum_{s \in \mathcal{S}_i}^{} P_s^t+\sum_{g \in \mathcal{G}_i}^{}\color{black}P_g^t\color{black} =\sum_{d \in \mathcal{D}_i}^{}P_d^t+(G_{ii}+\sum_{j\in\delta_i}^{}G_{ij}){V_i^t}^2+\nonumber\\
&\hspace{3cm}\sum_{j \in \delta_i}^{}p_{ij}^t+\sum_{e\in\mathcal{E}_i}^{}I_{e,t}^c\sum_{i \in \mathcal{I}_e}^{} V_i^t\label{ACOPF_lambda_i_p}\\
&\sum_{g \in \mathcal{G}_i}^{}\color{black}Q_g^t\color{black}-\sum_{d \in \mathcal{D}_i}^{}q_d^t=-(B_{ii}+\sum_{j\in \delta_i}^{}B_{ij}){V_i^t}^2+\sum_{j \in \delta_i}^{}q_{ij}^t\label{ACOPF_lambda_i_q}\\
&p_{ij}^t=-G_{ij}{V_i^t}^2+G_{ij}V_i^tV_j^t\cos{\theta_{ij}^t}-B_{ij}V_i^tV_j^t\sin{\theta_{ij}^t}\label{ACOPF_p}\\
&q_{ij}^t=B_{ij}(V_i^t)^2-B_{ij}V_i^tV_j^t\cos{\theta_{ij}^t}-G_{ij}V_i^tV_j^t\sin{\theta_{ij}^t}\label{ACOPF_q}\\ 
&\underline{V}_i \leq  V_i^t\leq \overline{V}_i \label{ACOPF_limv}\\
& \eqref{SOCP_mu_Ee}, \eqref{SOCP_mu_s}, \eqref{SOCP_lim}
\end{alignat}
\end{subequations} 

 The presented problem in \eqref{ACOPF} is a nonlinear optimization problem with PVs and EVs as DERs connected to the distribution network. Thus, the problem presented in \eqref{ACOPF} should be solved with nonlinear solvers such as IPOPT \cite{wachter2006implementation}.  

\vspace{-0.3cm}
\section{Case Studies}
\begin{figure}[h!]
    \centering
    \includegraphics[width=9.0cm]{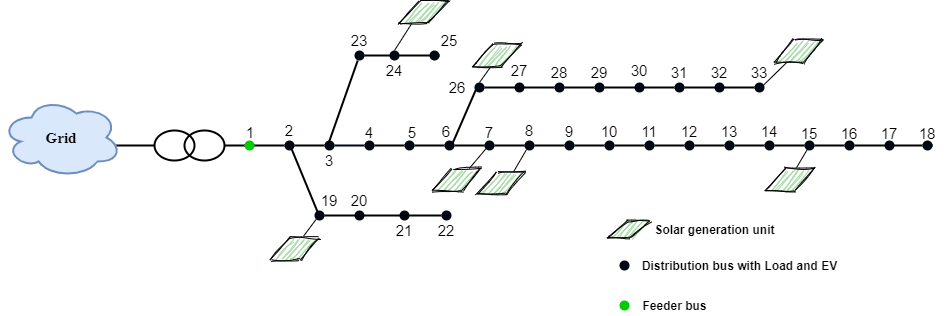}
    \vspace{-0.7cm}
    \caption{Modified IEEE 33-bus test system}
    \label{33-bus}
\end{figure}

In this section, the modified IEEE-33 bus system is leveraged to investigate the effect of utilizing high penetration EV on the voltage profile of the distribution network. The modified IEEE 33-bus system is presented in Fig. \ref{33-bus}. An EV charging station  is connected to each bus except the feeder bus of the modified IEEE 33-bus distribution network as shown in Fig. \ref{33-bus}. The base demand is set according to the normalized hourly load as well as solar generation data of California ISO on August 18, 2020\color{black}\cite{CAISO}\color{black}. 
\vspace{-0.5cm}
\vspace{-0.6cm}
\subsection{Comparing the Fixed Power and Fixed Current Models}\label{models}

\begin{figure}[h]
 \vspace{-0.4cm}
    \centering
    \includegraphics[width=9cm,height=5cm]{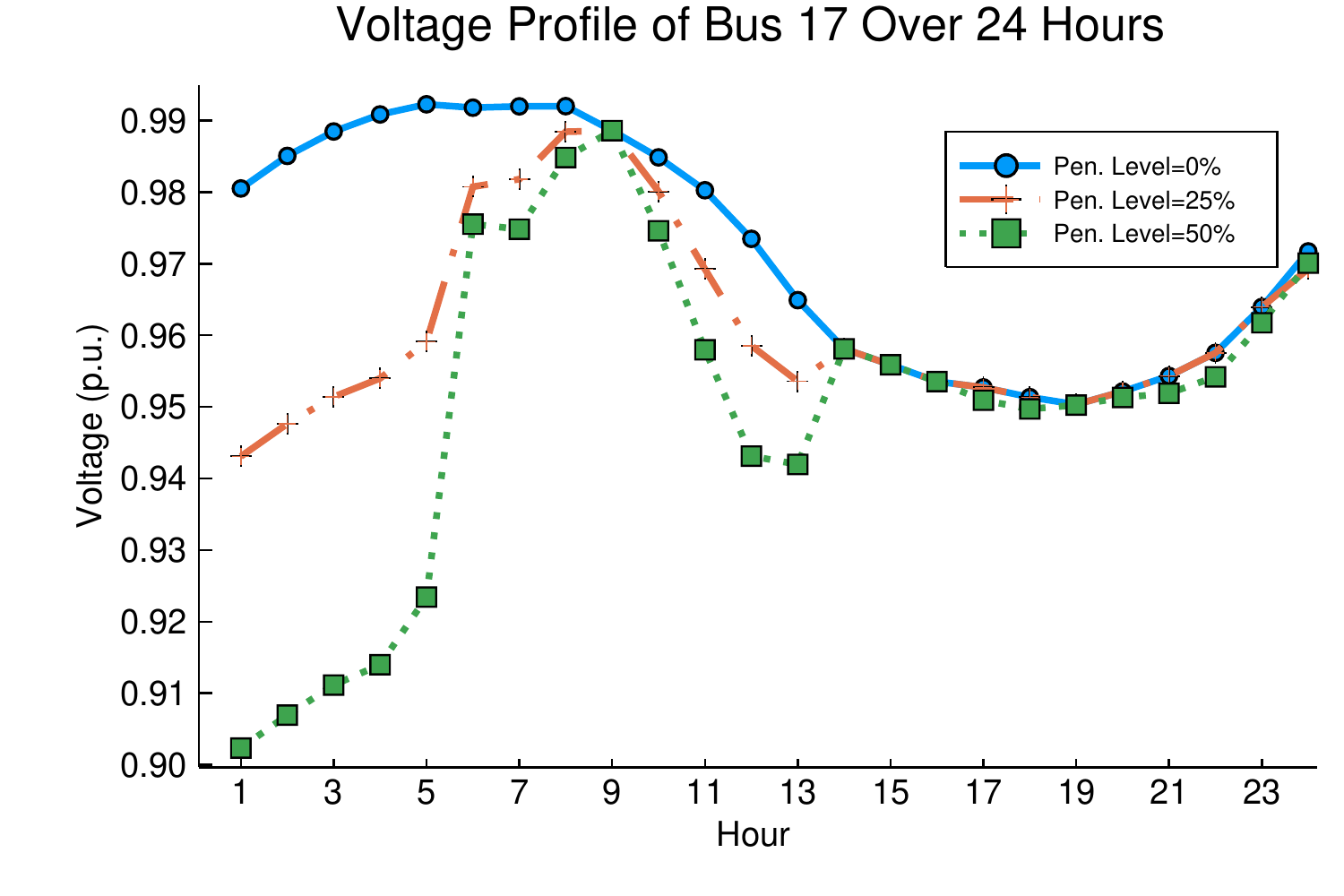}
    \vspace{-0.9cm}
    \caption{The comparison of voltage profile of bus 17 in 24 hours for different penetration levels}
    \vspace{-0.25cm}
    \label{bus_17_p_fixed}
    \vspace{-0.2cm}
\end{figure}

\begin{figure}[h!]
    \centering
    \includegraphics[width=9cm,height=5cm]{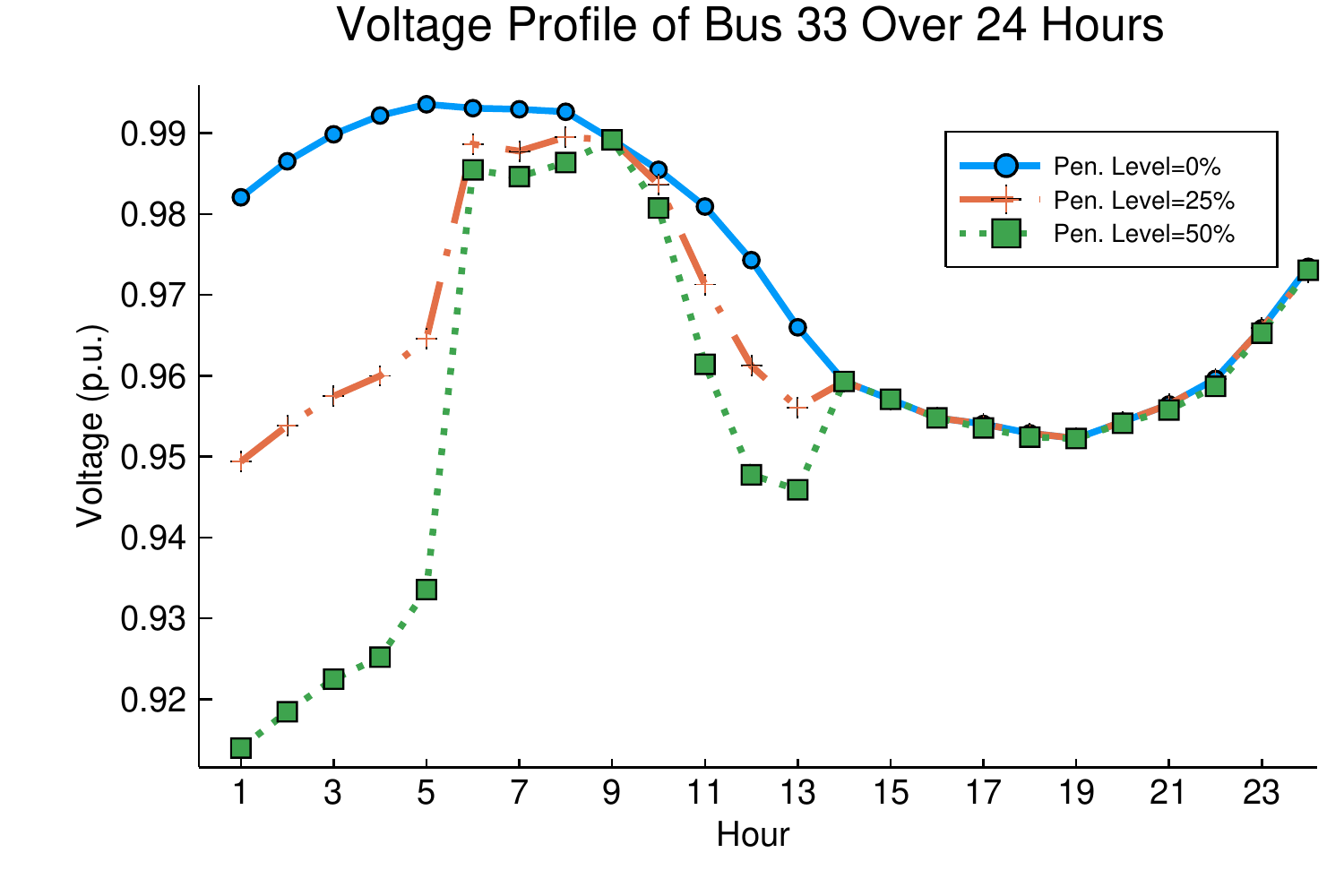}
    \vspace{-0.9cm}
    \caption{The comparison of voltage profile of bus 33 in 24 hours for different penetration levels}
    \vspace{-0.3cm}
    \label{bus_33_p_fixed}
    \vspace{-0.2cm}
\end{figure}

In Figs. \ref{bus_17_p_fixed} and \ref{bus_33_p_fixed}, the voltage profiles of \color{black} buses \color{black} $17$ and $33$ of the IEEE 33-bus system for different scenarios are presented, respectively. \color{black} Since these buses are far from the feeder, they were picked to show the worst voltage profiles in the distribution network. \color{black} Here, the charging power of EV is constant when the voltage magnitude varies. Figs. \ref{bus_17_p_fixed} and \ref{bus_33_p_fixed} show that when the penetration level of EV is increased, the voltage magnitude of some buses dropped under $0.95$ p.u. during some hours. For instance\color{black}, \color{black} the voltage of bus $17$ at hour $1$ is $0.9$ p.u. when the penetration level of EVs is increased to $50\%$ as shown in Fig. \ref{bus_17_p_fixed}. 

\begin{figure}[h!]
    \vspace{-0.4cm}
    \centering
    \includegraphics[width=9cm,height=5cm]{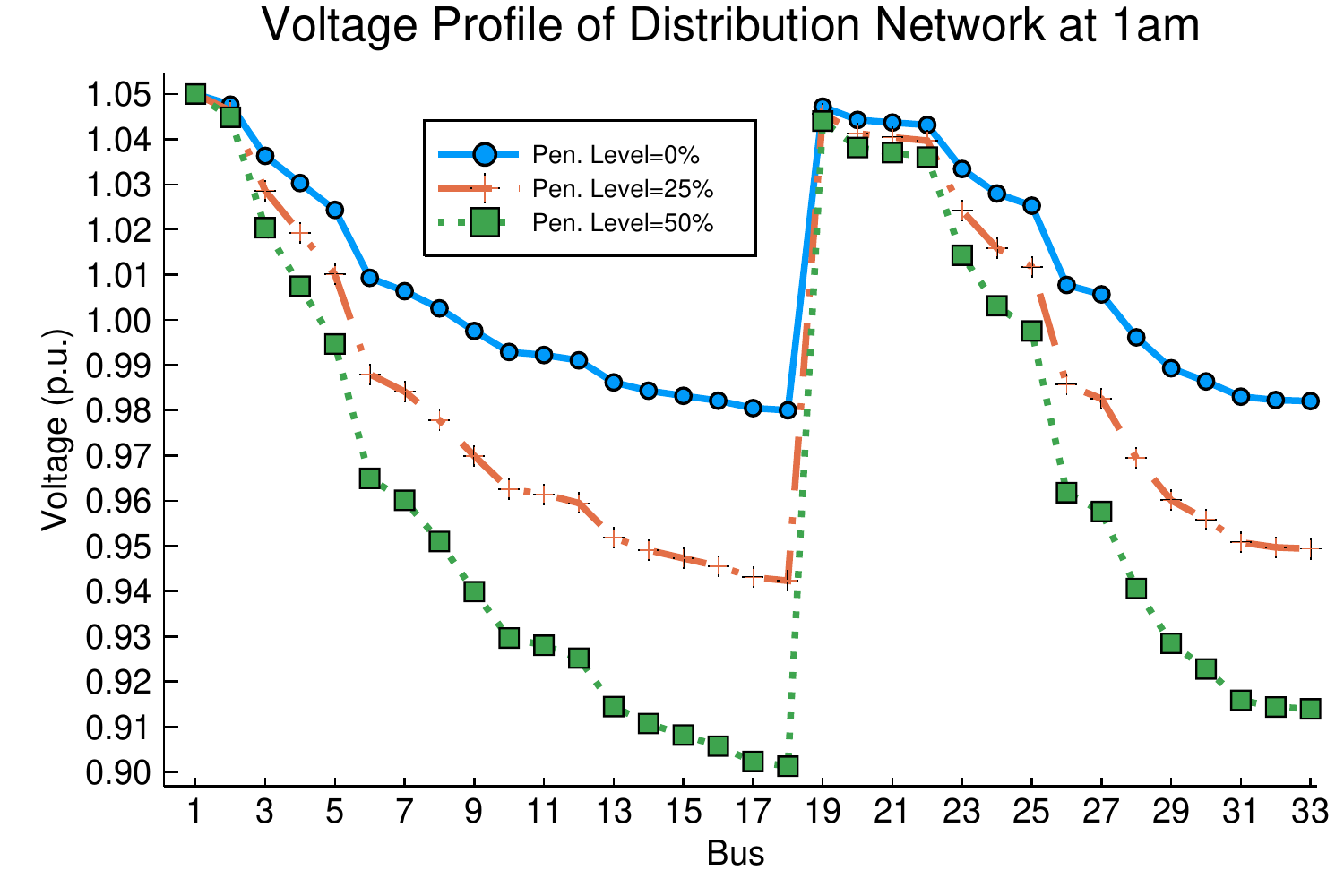}
    \vspace{-0.85cm}
    \caption{The comparison of voltage profile of buses at the first hour for different penetration levels}
    \label{time_1_p_fixed}
    \vspace{-0.45cm}
\end{figure}

Figs. \ref{bus_17_p_fixed} and \ref{bus_33_p_fixed} show that when the price of the electricity is low especially at night and noon, EVs tend to charge. Thus, the voltage magnitude of buses will drop more during those hours. To illustrate the voltage magnitude drop of buses, Fig. \ref{time_1_p_fixed} compares the voltage profile of buses at $1$am. Fig. \ref{time_1_p_fixed} illustrates that when the penetration level of EVs is zero the voltage magnitude of all buses is between $0.95$ and $1.05$ p.u. Increasing the penetration level of EVs leads to a decrease in voltage magnitude of buses. This decrease is more for those buses that are farther from the feeder bus for instance\color{black}, \color{black} buses $17$ and $33$.

\begin{figure}[h!]
    \vspace{-0.45cm}
    \centering
    \includegraphics[width=9cm,height=5cm]{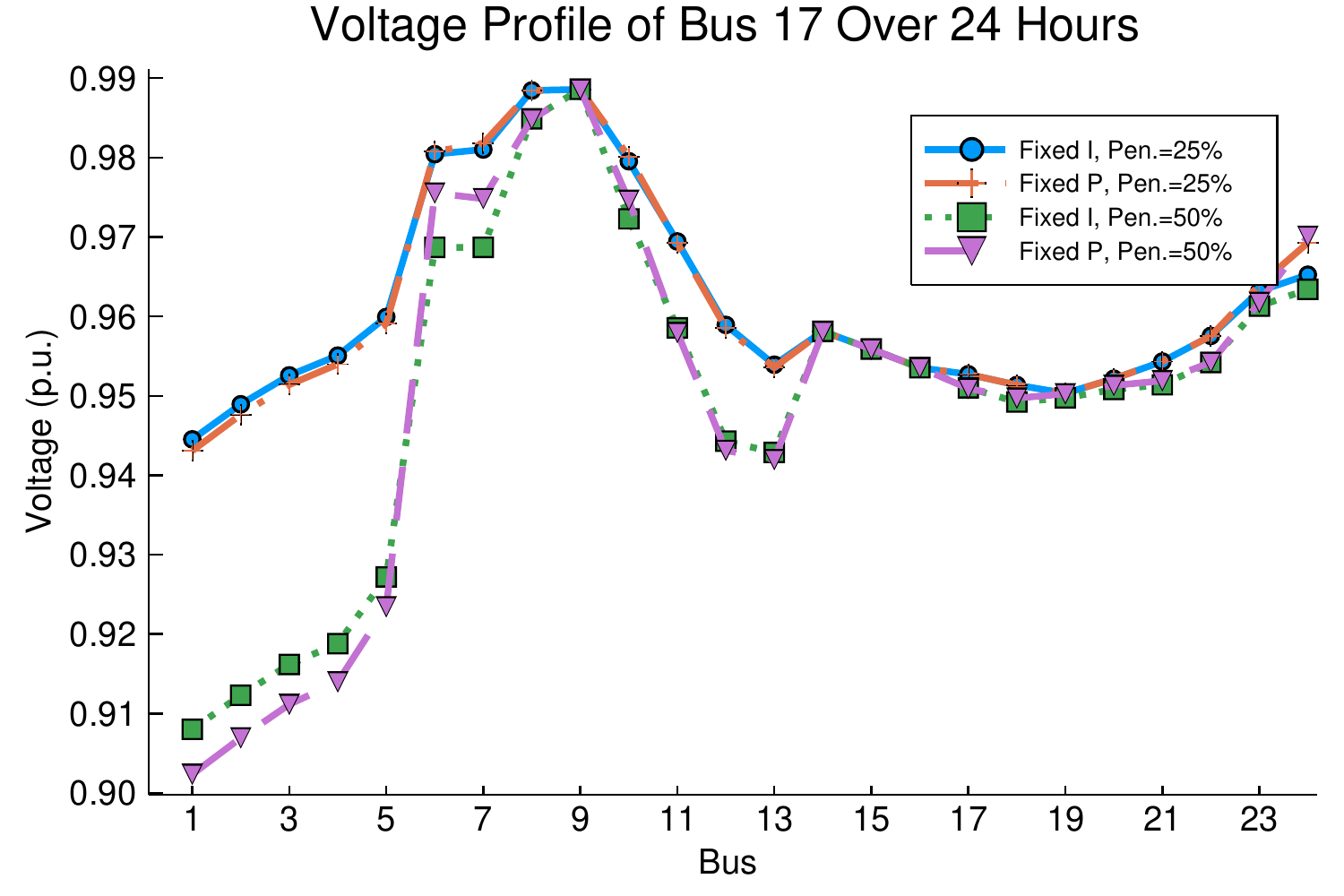}
    \vspace{-0.8cm}
    \caption{The comparison between leveraging EVs with fixed current and leveraging EVs with fixed power for bus $17$}
    \label{bus_17_comp}
    \vspace{-0.3cm}
\end{figure}

\begin{figure}[h!]
    \centering
    \includegraphics[width=9cm,height=5cm]{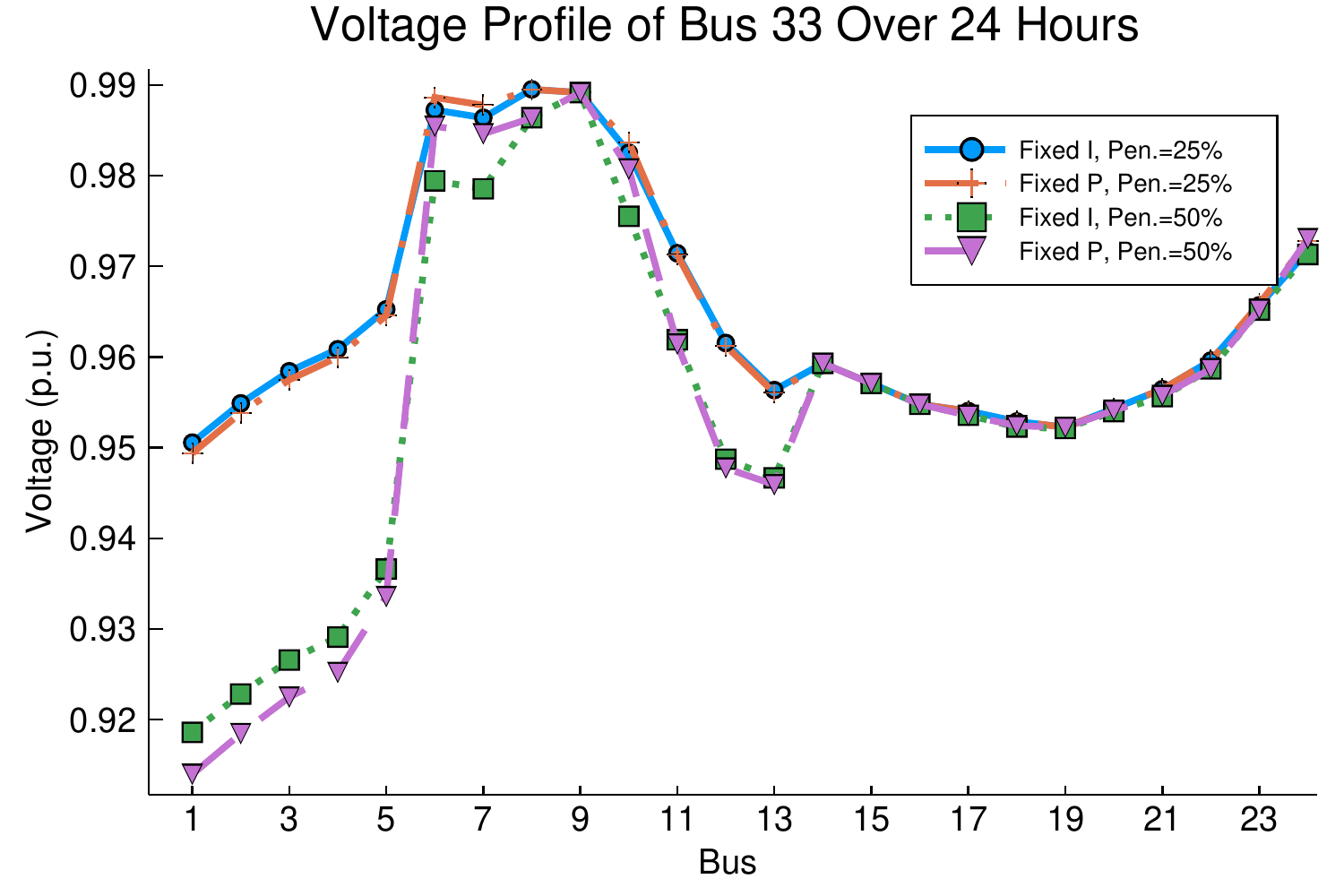}
    \vspace{-0.7cm}
    \caption{The comparison between leveraging EVs with fixed current and leveraging EVs with fixed power for bus $33$}
    \label{bus_33_comp}
    \vspace{-0.3cm}
\end{figure}

\begin{figure}[h!]
    \centering
    \includegraphics[width=9cm,height=5cm]{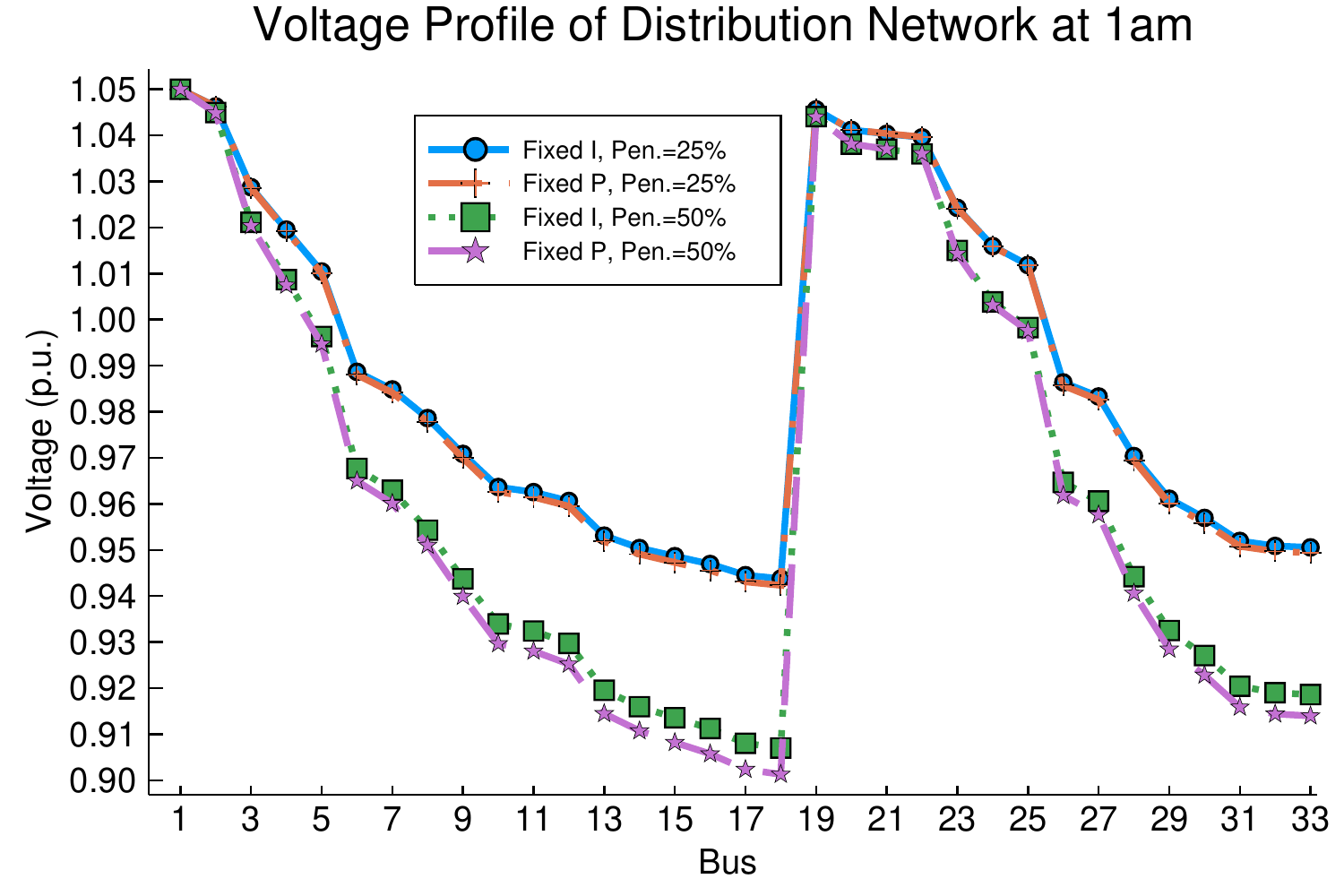}
    \vspace{-0.7cm}
    \caption{The comparison between leveraging EVs with fixed current and leveraging EVs with fixed power at $1$am}
    \label{time_1_comp}
    \vspace{-0.6cm}
\end{figure}

Since the EVs with fixed current receive less power when the voltage magnitude of their connected bus decreased, using the EVs with fixed current would mitigate the voltage magnitude drop. To compare the voltage profile procured by the fixed power model of EVs and the one procured by the fixed current model of EVs, Figs. \ref{bus_17_comp}, \ref{bus_33_comp}, and \ref{time_1_comp} are presented. Figs. \ref{bus_17_comp} and \ref{bus_33_comp}, illustrate the voltage profile of bus $17$ and $33$ over $24$ hours for different penetration levels and different model of EV. The voltage of buses $17$ and $33$ of the distribution network is increased when the fixed current model of EVs is employed in comparison with those procured by the fixed power model of EVs. However, the voltage magnitude of buses $17$ and $33$ procured by the fixed current model is less than $0.95$ p.u. during super off-peak hours. Fig. \ref{time_1_comp} presents the comparison between the voltage profile of buses procured by utilizing EVs with fixed current and the one procured by using EVs with fixed power at hour $1$ which illustrates a similar pattern with slight improvements in the voltage profile. Although, employing the fixed current model present some minor improvements in the voltage profile, with the increase to larger penetration levels, the voltage profile will fall under 0.95 p.u. 

\subsection{Sensitivity of Voltage Profile to Electricity Prices}
\begin{figure}[h!]
    \vspace{-0.3cm}
    \centering
    \includegraphics[width=9cm]{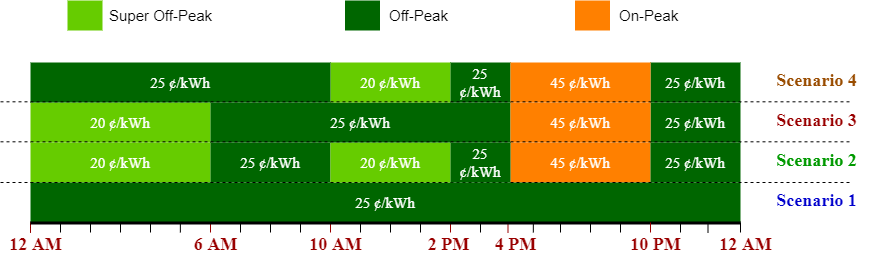}
    \vspace{-0.7cm}
    \caption{The TOU pricing for different scenarios}
    \label{TOU}
    \vspace{-0.4cm}
\end{figure}

\vspace{-0.1cm}
In this subsection, the impact of employing different TOU prices on the voltage profile of buses procured by the SOCP problem formulation presented in \eqref{SOCP} with a fixed power model of EVs is investigated. Note that the penetration level of all scenarios is $50\%$. \color{black} The TOU pricing for several \color{black} utilized scenarios are shown in Fig. \ref{TOU}.

\begin{figure}[h!]
    \vspace{-0.35cm}
    \centering
    \includegraphics[width=9cm,height=5cm]{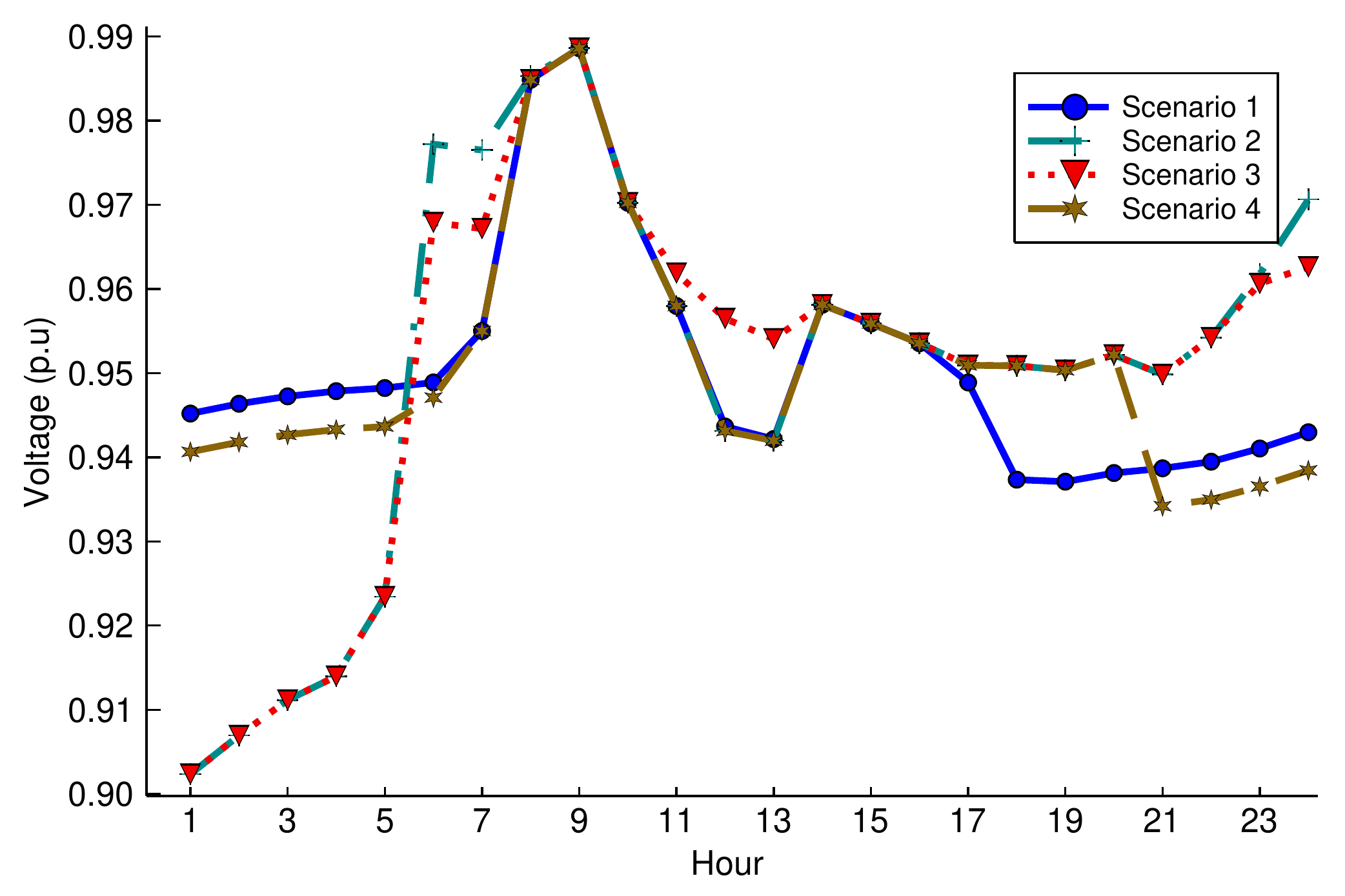}
    \vspace{-0.8cm}
    \caption{The sensitivity of voltage profile of bus $17$ on electricity prices}
    \label{bus_17_comp_TOU}
    \vspace{-0.5cm}
\end{figure}

\begin{figure}[h!]
    \centering
    \includegraphics[width=9cm,height=5cm]{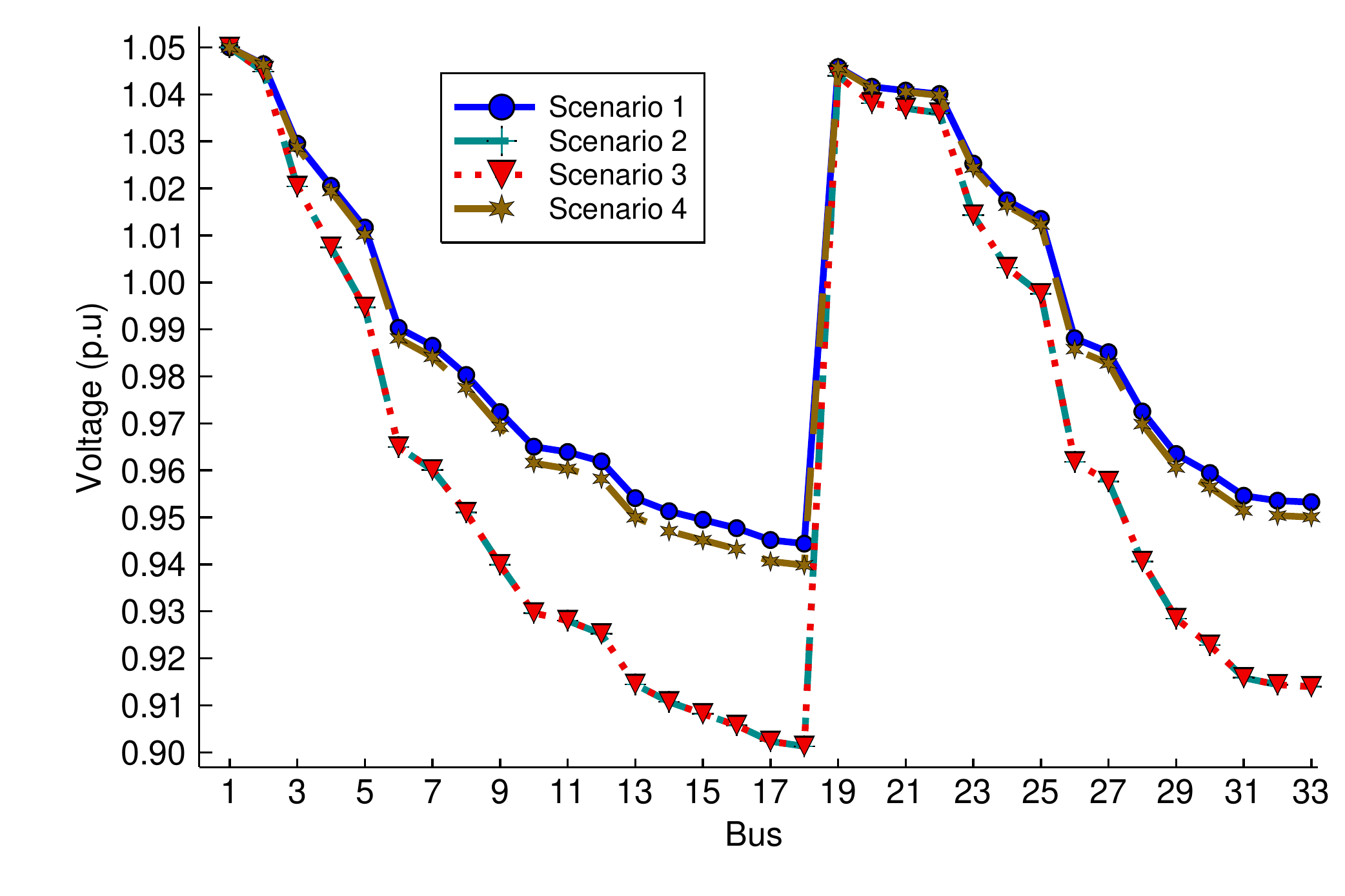}
    \vspace{-1cm}
    \caption{The sensitivity of voltage magnitude of all buses at hour $1$ on electricity prices}
    \label{time_1_comp_TOU}
    \vspace{-0.6cm}
\end{figure}

Fig. \ref{bus_17_comp_TOU} presents the voltage profile of buses $17$ over $24$ hours for different scenarios. The voltage magnitude of buses drops due to increasing charging power demand of EVs. It is shown that when the price of electricity is constant during $24$ hours, the voltage magnitude drop of bus $17$ is no longer an issue during the night. However, the voltage magnitude of bus $17$ for $18 \leq t \leq 24$ drops below the desired 0.95 p.u limit in this scenario. However, when the TOU pricing changes over 24 hours, the voltage magnitude of buses decreases in supper off-peak hours as presented for scenarios $2-4$ in Fig. \ref{bus_17_comp_TOU}. Fig. \ref{time_1_comp_TOU} shows the voltage magnitude of all buses at hour $1$ for different scenarios. Fig. \ref{time_1_comp_TOU} illustrates that the voltage magnitude of all buses at hour $1$ is increased for scenarios $1$ and $4$. The voltage magnitude of buses in scenarios 2 and 3 are the same at hour $1$. However, the voltage magnitude of bus $17$ in third scenario at $12\leq t\leq 13$ is less than that in second scenario. This is because of the fact that the TOU price of second scenario at hours $12,13$ is less than that for third scenario. Thus, the model tends to increase the charging power of EVs at hours $12,13$ and decrease the charging power of EVs at hours $6,7$ to minimize the operation cost. It should be noted that the TOU pricing utilized in section \ref{models} is scenario $2$.

\vspace{-0.3cm}

\section{Conclusion}
This paper investigated the impact of high penetration level of EVs on the voltage profile of distribution networks. In this paper, first, the SOCP relaxed form of the ACOPF problem of distribution network with solar generation and fixed power EVs is presented. Then the full ACOPF problem formulation of distribution network with solar generation and fixed current EVs is presented. The results show that employing the fixed current model of EVs can slightly decrease the voltage drop of buses of the distribution network. However, the voltage drop of buses is still a major issue when the fixed current model of EVs are employed. Then, the sensitivity of the voltage profile of buses on time-of-use pricing is investigated. Results suggest that such changes might be able to shift the voltage drop issue or mitigate it for certain hours of operation but it is not capable of entirely addressing the issue. \color{black} It should be noted that the model of \color{black} most of current chargers is the same as the fixed current model presented in this paper.
\vspace{-0.3cm}

\bibliographystyle{IEEEtran}
\bibliography{mend.bib}
\end{document}